\documentclass[preprint,aps,12pt,showpacs,nofootinbib,tightenlines]{revtex4}
\usepackage{amsmath}
\usepackage{amssymb}
\usepackage{epsfig}
\usepackage{graphicx}
\begin{document}
\def\pslash{\rlap{\hspace{0.02cm}/}{p}}
\def\eslash{\rlap{\hspace{0.02cm}/}{e}}
\title {Pair production of the heavy leptons associated with a gauge boson $\gamma$ or $Z$ at the ILC }
\author{ Qing-Guo Zeng}
\author{ Chong-Xing Yue}\email{cxyue@lnnu.edu.cn}
\author{ Jiao Zhang}
\affiliation{\small Department of Physics, Liaoning Normal University, Dalian
116029, China \\
     \vspace*{1.5cm}}
\begin{abstract}
The T-odd leptons are the typical particles predicted by the littlest Higgs model with T-parity (LHT model) and the
observation of these particles might be regarded as the direct evidence of the LHT model. In this paper, we investigate
the production of a pair of
T-odd leptons associated with a gauge boson $V$($\gamma$ or $Z$)  at the international
linear $e^{+}e^{-}$ collider (ILC). The numerical results show that the
possible signals of the T-odd leptons
may be detected in the future ILC experiments.
\end{abstract}
\pacs{14.60.Hi, 12.60.-i, 13.66.De} \maketitle
\section{ Introduction}
\noindent

 The standard model (SM) provides an excellent effective field theory description of
 almost all particle physics experiments.
 However, the theoretical shortcomings of the SM, such as quadratic divergencies, the
 triviality of a $\phi^{4}$ theory, etc,
 suggest that it should be embedded in a larger scheme.
 Many popular new physics (NP) models beyond the SM have been proposed, and some of
 which predict the existence of new charged
 leptons. Any signal for such kind of particles in future high energy experiments will
 play a milestone role in discovery
 of NP. Thus, studying production and decay of the new charged leptons in future high
 energy collider experiments is of special interest.

Little Higgs theory \cite{lh4} is proposed as an interesting solution to the so-called hierarchy problem of the SM and
can be regarded as one of the important candidates for NP beyond the SM. Among of
the little Higgs models, the littlest
Higgs (LH) model \cite{lh5} has all essential features of the little Higgs models.
However, the original version of the
LH model suffers from precision electroweak (EW) constrains, the NP effects are
small as the NP scale $f$ is required to
be above 2-3$TeV$ in order to satisfy the EW precision constraints, which re-introduces
the fine tuning and the little
hierarchy problem \cite{lh5}\cite{lhf}.
The LH model with T-parity (LHT) \cite{lht1}\cite{lht2}\cite{lht3} is one of the
attractive little Higgs models.
In the LHT model, all dangerous tree-level contributions to low energy EW observables
are forbidden by T-parity and
hence the corrections to low energy EW observables are loop-suppressed and
small \cite{lht1}\cite{flv9-11}. As a result,
the relatively low new particle mass scale $f$ is still allowed by data,
e.g., $f>500GeV$ \cite{flv9-11}.

In the LHT model, particle fields are divided into T-even  and T-odd sectors under
 T-parity and the SM fields are T-even.
In order to implement T-parity in the fermion sector, one introduces three doublets of
mirror quarks and mirror leptons,
 which have T-odd parity, transform vectorially under $SU(2)_{L}$ and can be given
 large masses. These mirror fermions  have
 new flavor
violating interactions with the SM fermions mediated by the new gauge bosons and at
 higher order by the triplet scalar, which might
generate significantly contributions to some flavor violation processes \cite{lht3}\cite{flv9-11}\cite{flv12}\cite{flv13-14}.
It has been shown that the LHT mirror fermion interactions can yield large NP effects in the quark sector \cite{lht3}\cite{flv9-11}\cite{flv13-14} and
the lepton sector \cite{12}\cite{219}\cite{c219}.

So far, lots of studies about the heavy charged leptons have previously been done
at hadron colliders. However, there exists some difficulties to detect heavy lepton production at the CERN large hadron
 collider (LHC), due to large backgrounds. Compared to hadron colliders, the future international linear $e^{+}e^{-}$ collider (ILC)
has the advantage in performing
experimental measurement with a particularly clean environment \cite{42647}. Furthermore, ILC can provide complementary information for
NP with performing precision measurements that would complete the LHC results. Many works involving the new charged leptons
have been given at the ILC.  Studies about the heavy charged leptons predicted by the LHT model have previously been done at the LHC \cite{71} and ILC\cite{72}.
As a complementary production mode to the former research, this paper is to study the
production processes of the T-odd leptons in association with a neutral gauge boson $V(=\gamma$, or $Z$) at the ILC experiments.

This paper is organized as follows. In section II, we give a brief review of the LHT model and then give the
 relevant couplings. In section III, we devote to the computation of the production cross section of
 the process $e^+e^-\rightarrow \gamma\overline{L}_{i}{L}_{j}$. The study of the production process
 $e^+e^-\rightarrow Z\overline{L}_{i}{L}_{j}$ is presented in section IV. Some phenomenological
 analysis are included in the above two sections. The conclusions are given in section V.

\section{ A brief review of the LHT model}
 \noindent

The LHT model is based on an $SU(5)/SO(5)$ global symmetry breaking pattern. A subgroup
$[SU(2)\times U(1)]_{1}\times[SU(2)\times U(1)]_{2}$ of the $SU(5)$ global symmetry is gauged,
and at the scale $f$ it is broken into the SM EW symmetry $SU(2)_{L}\times U(1)_{Y}$. T-parity is an
automorphism that exchanges the $[SU(2)\times U(1)]_{1}$ and $[SU(2)\times U(1)]_{2}$ gauge symmetries.
The T-even combinations of the gauge fields are the EW gauge bosons, and the T-odd combinations are their
T-parity partners. After taking into account EW symmetry breaking, at the order of $\nu^{2}/f^{2}$,
the masses of the T-odd set of the $SU(2)\times U(1)$ gauge bosons are given by:
\begin {equation}
M_{Z_{H}}=M_{W_{H}}=gf(1-\frac{\nu ^{2}}{8f^{2}}),~~~M_{B_{H}}=\frac{g'f}{\sqrt{5}}
(1-\frac{5\nu ^{2}}{8f^{2}}).
\end {equation}
Where $g$ and $g'$ are the corresponding coupling constants of $SU(2)_{L}$ and $U(1)_{Y}$.
 $ \nu = 246 GeV $ is the EW
scale and $f$ is the
scale parameter of the gauge symmetry breaking of the LHT model. Moreover, because of the
 smallness of $g'$, the T-odd gauge boson ${B}_{H}$ is the lightest T-odd particle, which is stable,
 electrically neutral, and weakly interacting particle. Thus, it can be seen as an attractive dark
 matter candidate \cite{10}. To avoid severe constraints and simultaneously implement T-parity,
 one needs to double the SM fermion doublet spectrum  \cite{lht1}\cite{lht2}. The T-even combination is
 associated with the $SU(2)_{L}$ doublet, while the T-odd combination is its T-parity partner.
  At the leading order, the masses of the T-odd fermions can be written in a unified manner as:
\begin{equation}
M_{F_{H}^{i}}=\sqrt{2}\kappa_if
\end{equation}
 where the Yukawa couplings $\kappa_i$ can in general depend on the fermion species $i$.

  One of the important ingredients of the mirror sector in the LHT model is the existence
  of CKM-like unitary mixing matrices. Mirror fermions are characterized by new flavor
  interactions with SM fermions and heavy gauge bosons, which involve two new unitary
  mixing matrices in the quark sector, $V_{Hu}$ and $V_{Hd}$, and two in the lepton
  sector, $V_{Hl}$ and $V_{H\nu}$ \cite{flv9-11}\cite{flv12}.
 These mirror mixing matrices parameterize
  flavor-changing (FC) interactions between the SM fermions and the mirror fermions.
 The two CKM-like unitary mixing matrices  $V_{Hl}$ and $V_{H\nu}$ satisfy the following physical constrains:
\begin{equation}
V^{\dag}_{H\nu}V_{Hl}=V_{PMNS}.
\end{equation}
Here the Pontecorvo-Maki-Nakagata-Saki (PMNS) matrix $V_{PMNS}$ is defined through neutrino mixing.
 $V_{Hl}$, the most important mixing matrix in the present paper, parameterizes the interactions of light charged leptons with mirror neutrinos, mediated by $W^{\pm}_{H}$, and with mirror charged leptons, mediated by $Z_{H}$ and $B_{H}$.  On the other hand, $V_{H\nu}$ parameterizes the interactions of light neutrinos with mirror leptons.
 Ref. \cite{lht3} parameterizes $V_{Hl}$ with three mixing angles $\theta^l_{12},\theta^l_{23},\theta^l_{13}$ and three complex phases
$\delta^l_{12},\delta^l_{23},\delta^l_{13}$:
\begin{eqnarray}
V_{Hl}=
\begin{pmatrix}
c^l_{12}c^l_{13}&s^l_{12}c^l_{13}e^{-i\delta^l_{12}}&s^l_{13}e^{-i\delta^l_{13}}\\
-s^l_{12}c^l_{23}e^{i\delta^l_{12}}-c^l_{12}s^l_{23}s^l_{13}e^{i(\delta^l_{13}-\delta^l_{23})}&
c^l_{12}c^l_{23}-s^l_{12}s^l_{23}s^l_{13}e^{i(\delta^l_{13}-\delta^l_{12}-\delta^l_{23})}&
s^l_{23}c^l_{13}e^{-i\delta^l_{23}}\\
s^l_{12}s^l_{23}e^{i(\delta^l_{12}+\delta^l_{23})}-c^l_{12}c^l_{23}s^l_{13}e^{i\delta^l_{13}}&
-c^l_{12}s^l_{23}e^{i\delta^l_{23}}-s^l_{12}c^l_{23}s^l_{13}e^{i(\delta^l_{13}-\delta^l_{12})}&
c^l_{23}c^l_{13}
\end{pmatrix}.
\end{eqnarray}
For the matrix $V_{PMNS}$, we take the standard parameterization form with parameters given by the neutrino
experiments \cite{epc59}. As no constraints on the PMNS phases exist, we will set the three Majorana phases of $V_{PMNS}$ to  zero in our numerical estimations.

The Feynman rules of the T-odd leptons (mirror leptons) which are related to our calculation can be written as \cite{lht3}:
\begin{eqnarray}
\gamma\overline{L}_{i}{L}_{j}:-ie\gamma^{{\mu}}\delta_{ij}~,~~~~~~~~~~~~~~~~~
{B}_{H}\overline{L}_{i}{l}_{j}:\frac{ie}{C_W}[\frac{1}{10}+\frac{5C^2_W}{8(5C^2_W-S^2_W)}\frac{v^2}{f^2}]
(V_{Hl})_{ij}\gamma^{{\mu}}P_L;~~~~\\
Z\overline{L}_{i}{L}_{j}:\frac{ie}{S_WC_W}[-\frac{1}{2}+S^2_W]\gamma^{{\mu}}\delta_{ij}~;~~
{Z}_{H}\overline{L}_{i}{l}_{j}:\frac{ie}{S_W}[-\frac{1}{2}+\frac{S^2_W}{8(5C^2_W-S^2_W)}\frac{v^2}{f^2}]
(V_{Hl})_{ij}\gamma^{{\mu}}P_L
\end{eqnarray}
where $P_L=\frac{1}{2}(1-\gamma_5)$ is the left-handed projection operator. $S_{W}$ represents the ${\sin\theta_{W}}$ of the Weinberg angle $\theta_{W}$. {${l}_{i}$} and {${L}_{j}$} represent the three family leptons $e$, $\mu$, and
$\tau$ and the three family T-odd leptons, respectively.

Certainly, the trilinear coupling ${Z}_{H}{B}_{H}Z$ can also contribute to the process $e^+e^-\rightarrow V\overline{L}_{i}{L}_{j}$. However, this kind of couplings are induced at the one-loop level by a fermion triangle and its contributions are very small \cite{6206}. Thus, in our following calculation, we will neglect the contributions of this kind of couplings .

\begin{figure}[htbp]
\includegraphics [scale=0.8] {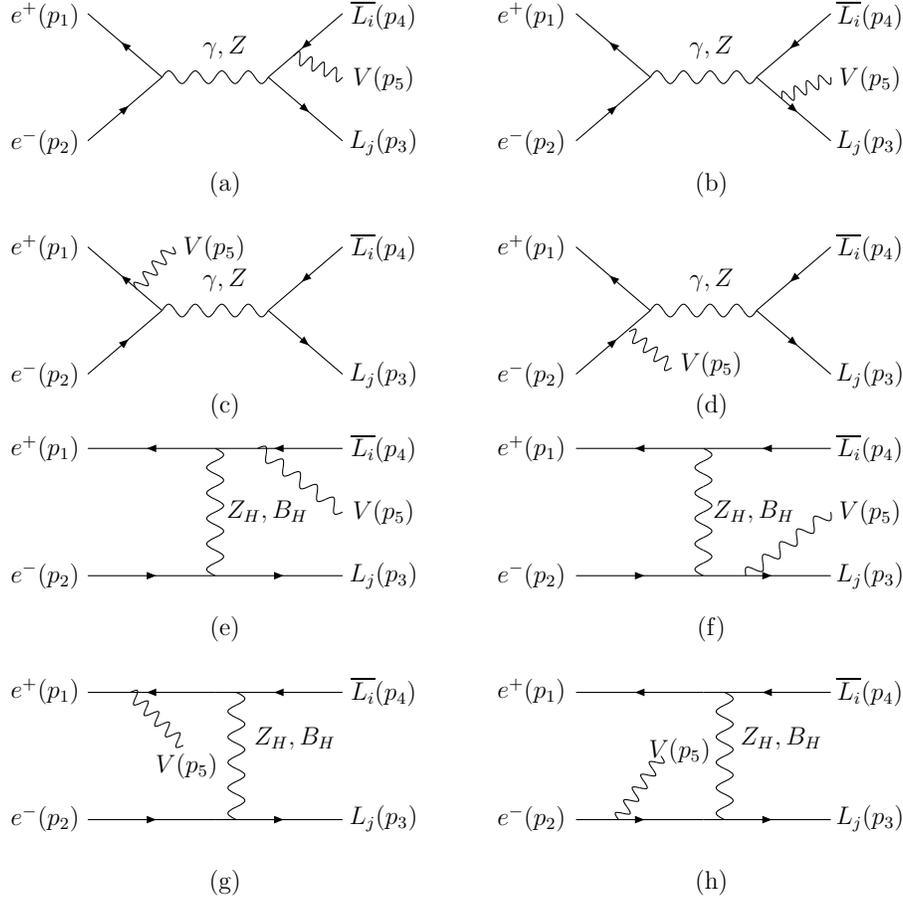}
\caption{The Feynman diagrams of the process $e^+e^-\rightarrow V\overline{L}_{i}{L}_{j}$ in the LHT model. }
\end{figure}
\vspace{-0.5cm}

\section{the processes $e^+e^-\rightarrow \gamma\overline{L}_{i}{L}_{j}$}
\noindent

With the above couplings, the Feynman diagrams for the process $e^+(p_{1})e^-(p_{2})\rightarrow \gamma(p_5)\overline{L}_{i}(p_{4}){L}_{j}(p_{3})$ are shown in Fig.1. The production  amplitude  can be written as
\begin{eqnarray}
{M}_{1}&=&{M}^{\gamma\gamma}_{a}+{M}^{Z\gamma}_{a}+{M}^{\gamma\gamma}_{b}+{M}^{Z\gamma}_{b}
+{M}^{\gamma\gamma}_{c}+{M}^{Z\gamma}_{c}+{M}^{\gamma\gamma}_{d}+{M}^{Z\gamma}_{d}
+{M}^{B_{H}\gamma}_{e}\nonumber\\&&
+{M}^{Z_{H}\gamma}_{e}+{M}^{B_{H}\gamma}_{f}+{M}^{Z_{H}\gamma}_{f}
+{M}^{B_{H}\gamma}_{g}+{M}^{Z_{H}\gamma}_{g}+{M}^{B_{H}\gamma}_{h}+{M}^{Z_{H}\gamma}_{h}
\end{eqnarray}
with
\begin{eqnarray}
 {M}^{\gamma\gamma}_{a}&=&-ie^3G(p_1+p_2,0)G(p_4+p_5,{M}_{L})\bar{v}(p_{1})\gamma^{\mu}u(p_{2})
 \bar{u}(p_{3})\gamma_{\mu}[-(\pslash_4+\pslash_5)+{M}_{L}]\nonumber\\&&\times\rlap/\epsilon(p_5)v(p_{4}),
\end{eqnarray}
\vspace{-1.cm}
\begin{eqnarray}
 {M}^{Z\gamma}_{a}&=&\frac{-ie^3}{S^2_WC^2_W}(-\frac{1}{2}+S^2_W)G(p_1+p_2,{M}_{Z})G(p_4+p_5,{M}_{L})\bar{v}(p_{1})\gamma^{\mu}
 [(-\frac{1}{2}+S^{2}_W)P_{L}\nonumber\\&&
 +(S^{2}_W)P_{R}]u(p_{2})\bar{u}(p_{3})\gamma_{\mu}[-(\pslash_4+\pslash_5)+{M}_{L}]\rlap/\epsilon(p_5)v(p_{4}),
\end{eqnarray}
 \begin{eqnarray}
 {M}^{\gamma\gamma}_{b}&=&-ie^3G(p_1+p_2,0)G(p_3+p_5,{M}_{L})\bar{v}(p_{1})
 \gamma^{\mu}u(p_{2})\bar{u}(p_{3})\rlap/\epsilon(p_5)[\pslash_3+\pslash_5+{M}_{L}]\nonumber\\&&\times\gamma_{\mu}v(p_{4}),
\end{eqnarray}
\begin{eqnarray}
 {M}^{Z\gamma}_{b}&=&\frac{-ie^3}{S^2_WC^2_W}(-\frac{1}{2}+S^2_W)G(p_1+p_2,{M}_{Z})G(p_3+p_5,{M}_{L})\bar{v}(p_{1})
 \gamma^{\mu}[(-\frac{1}{2}+S^{2}_W)P_{L}\nonumber\\&&+(S^{2}_W)P_{R}]
 u(p_{2})\bar{u}(p_{3})\rlap/\epsilon(p_5)[\pslash_3+\pslash_5+{M}_{L}]\gamma_{\mu}v(p_{4}),
\end{eqnarray}
\begin{eqnarray}
 {M}^{\gamma\gamma}_{c}&=&-ie^3G(p_1-p_5,{M}_{e})G(p_3+p_4,0)\bar{v}(p_{1})\rlap/\epsilon(p_5)
 [-(\pslash_1-\pslash_5)+{M}_{e}]\gamma^{\mu}u(p_{2})\bar{u}(p_{3})\nonumber\\&&\times\gamma_{\mu}v(p_{4}),
\end{eqnarray}
\begin{eqnarray}
 {M}^{Z\gamma}_{c}&=&\frac{-ie^3}{S^2_WC^2_W}(-\frac{1}{2}+S^2_W)G(p_1-p_5,{M}_{e})G(p_3+p_4,{M}_{Z})\bar{v}(p_{1})
 \rlap/\epsilon(p_5)[-(\pslash_1-\pslash_5)\nonumber\\&&+{M}_{e}]\gamma^{\mu}
 [(-\frac{1}{2}+S^{2}_W)P_{L}+(S^{2}_W)P_{R}]u(p_{2})
 \bar{u}(p_{3})\gamma_{\mu}v(p_{4}),
\end{eqnarray}
\begin{eqnarray}
 {M}^{\gamma\gamma}_{d}&=&-ie^3G(p_2-p_5,{M}_{e})G(p_3+p_4,0)\bar{v}(p_{1})
 \gamma^{\mu}[(\pslash_2-\pslash_5)+{M}_{e}]\rlap/\epsilon(p_5)u(p_{2})\bar{u}(p_{3})\nonumber\\&&\times\gamma_{\mu} v(p_{4}),
 \end{eqnarray}
\begin{eqnarray}
 {M}^{Z\gamma}_{d}&=&\frac{-ie^3}{S^2_WC^2_W}(-\frac{1}{2}+S^2_W)G(p_2-p_5,{M}_{e})G(p_3+p_4,{M}_{Z})\bar{v}(p_{1})
 \gamma^{\mu}[(-\frac{1}{2}+S^{2}_W)P_{L}\nonumber\\&&+(S^{2}_W)P_{R}][(\pslash_2-\pslash_5)+{M}_{e}]\rlap/\epsilon(p_5)
 u(p_{2})\bar{u}(p_{3})\gamma_{\mu}v(p_{4}),
\end{eqnarray}
\begin{eqnarray}
 {M}^{{B}_{H}\gamma}_{e}&=&\frac{-ie^3}{C^2_W}[\frac{1}{10}+\frac{5C^2_W}{8(5C^2_W-S^2_W)}\frac{\nu^2}{f^2}]^2(V_{Hl})_{ie}(V_{Hl})_{ej}
 G(p_4+p_5,{M}_{L})G(p_2-p_3,{M}_{{B_{H}}})\nonumber\\&&\times\bar{v}(p_{1})\gamma^{\mu}P_{L}
 [-(\pslash_4+\pslash_5)+{M}_{L}]\rlap/\epsilon(p_5)v(p_{4})\bar{u}(p_{3})\gamma_{\mu}P_{L}u(p_{2}),
\end{eqnarray}
\begin{eqnarray}
 {M}^{{Z}_{H}\gamma}_{e}&=&\frac{-ie^3}{S^2_W}[-\frac{1}{2}+\frac{S^2_W}{8(5C^2_W-S^2_W)}\frac{\nu^2}{f^2}]^2(V_{Hl})_{ie}(V_{Hl})_{ej}
 G(p_4+p_5,{M}_{L})G(p_2-p_3,{M}_{{Z_{H}}})\nonumber\\&&\times\bar{v}(p_{1})\gamma^{\mu}P_{L}
 [-(\pslash_4+\pslash_5)+{M}_{L}]\rlap/\epsilon(p_5)v_(p_{4})\bar{u}(p_{3})\gamma_{\mu}P_{L}u_(p_{2}),
\end{eqnarray}
\begin{eqnarray}
 {M}^{{B}_{H}\gamma}_{f}&=&\frac{-ie^3}{C^2_W}[\frac{1}{10}+\frac{5C^2_W}{8(5C^2_W-S^2_W)}\frac{\nu^2}{f^2}]^2(V_{Hl})_{ie}(V_{Hl})_{ej}
 G(p_4-p_1,{M}_{{B}_{H}})G(p_3+p_5,{M}_{L})\nonumber\\&&\times\bar{v}(p_{1})\gamma^{\mu}P_{L}
  v(p_{4})\bar{u}(p_{3})\rlap/\epsilon(p_5)[\pslash_3+\pslash_5+{M}_{L}]\gamma_{\mu}P_{L}u(p_{2}),
\end{eqnarray}
\begin{eqnarray}
 {M}^{{Z}_{H}\gamma}_{f}&=&\frac{-ie^3}{S^2_W}[-\frac{1}{2}+\frac{S^2_W}{8(5C^2_W-S^2_W)}\frac{\nu^2}{f^2}]^2(V_{Hl})_{ie}(V_{Hl})_{ej}
 G(p_4-p_1,{M}_{{Z}_{H}})G(p_3+p_5,{M}_{L})\nonumber\\&&\times\bar{v}(p_{1})\gamma^{\mu}P_{L}
  v(p_{4})\bar{u}(p_{3})\rlap/\epsilon(p_5)[\pslash_3+\pslash_5+{M}_{L}]\gamma_{\mu}P_{L}u(p_{2}),
\end{eqnarray}
\begin{eqnarray}
 {M}^{{B}_{H}\gamma}_{g}&=&\frac{-ie^3}{C^2_W}[\frac{1}{10}+\frac{5C^2_W}{8(5C^2_W-S^2_W)}\frac{\nu^2}{f^2}]^2(V_{Hl})_{ie}(V_{Hl})_{ej}
 G(p_5-p_1,{M}_{e})G(p_2-p_3,{M}_{{B_{H}}})\nonumber\\&&\times\bar{v}(p_{1})\rlap/\epsilon(p_5)
 [\pslash_5-\pslash_1+{M}_{e}]\gamma^{\mu}P_{L}
 v(p_{4})\bar{u}(p_{3})\gamma_{\mu}P_{L}u(p_{2}),
\end{eqnarray}
\begin{eqnarray}
 {M}^{{Z}_{H}\gamma}_{g}&=&\frac{-ie^3}{S^2_W}[-\frac{1}{2}+\frac{S^2_W}{8(5C^2_W-S^2_W)}\frac{\nu^2}{f^2}]^2(V_{Hl})_{ie}(V_{Hl})_{ej}
 G(p_5-p_1,{M}_{e})G(p_2-p_3,{M}_{{Z_{H}}})\nonumber\\&&\times\bar{v}(p_{1})\rlap/\epsilon(p_5)
 [\pslash_5-\pslash_1+{M}_{e}]\gamma^{\mu}P_{L}
 v(p_{4})\bar{u}(p_{3})\gamma_{\mu}P_{L}u(p_{2}),
\end{eqnarray}
\begin{eqnarray}
 {M}^{{B}_{H}\gamma}_{h}&=&\frac{-ie^3}{C^2_W}[\frac{1}{10}+\frac{5C^2_W}{8(5C^2_W-S^2_W)}\frac{\nu^2}{f^2}]^2(V_{Hl})_{ie}(V_{Hl})_{ej}
 G(p_4-p_1,{M}_{{B}_{H}})G(p_2-p_5,{M}_{e})\nonumber\\&&\times\bar{v}(p_{1})\gamma^{\mu}P_{L}
  v(p_{4})\bar{u}(p_{3})\gamma_{\mu}P_{L}[\pslash_2-\pslash_5+{M}_{e}]\rlap/\epsilon(p_5)u(p_{2}),
\end{eqnarray}
\begin{eqnarray}
 {M}^{{Z}_{H}\gamma}_{h}&=&\frac{-ie^3}{S^2_W}[-\frac{1}{2}+\frac{S^2_W}{8(5C^2_W-S^2_W)}\frac{\nu^2}{f^2}]^2(V_{Hl})_{ie}(V_{Hl})_{ej}
 G(p_4-p_1,{M}_{{Z}_{H}})G(p_2-p_5,{M}_{e})\nonumber\\&&\times\bar{v}(p_{1})\gamma^{\mu}P_{L}
  v(p_{4})\bar{u}(p_{3})\gamma_{\mu}P_{L}[\pslash_2-\pslash_5+{M}_{e}]\rlap/\epsilon(p_5)u(p_{2}).
\end{eqnarray}
Where $G(p, M)=\frac{1}{p^2-M^2}$ denotes the propagator of the particle. $p_{1}$ and $p_{2}$ refer to the incoming momentum of the
incoming $e^+$ and $e^-$, respectively. $p_{4},   p_{3}$  and $ p_{5}$ are the momenta of the outgoing  final states {$\overline{L}_{i}$},
 {${L}_{j}$} and $\gamma$.

With the above production amplitudes, the production cross section can be directly obtained. In the calculation of the cross section,
instead of calculating the square of the amplitudes analytically, we calculate the amplitudes numerically by using the method of \cite{15},
 which can simplify our calculation. In our following calculation, the SM input parameters are taken as $S_{W}^{2}=0.231$,
 ${m}_{Z}=91.187GeV$ and the fine-structure constant $\alpha=1/128$ \cite{16}.

From above discussions, we can see that the production cross sections $ \sigma(\gamma\overline{L}_{i}{L}_{j})$  for the processes $e^+e^-\rightarrow \gamma\overline{L}_{i}{L}_{j}$
 are dependent on the model-dependent free parameters, the symmetry breaking scale $f$, the mirror lepton masses ${M}_{{L_{i}}}$,  and  the matrix elements ${(V_{Hl})}_{ij}$.  The matrix elements ${(V_{Hl})}_{ij}$ can be determined through $ V_{Hl}=V_{H\nu}V_{PMNS} $. In order to simply the calculation and avoid any additional parameters, we take  $ V_{Hl}=V_{PMNS} $, which means that the T-odd leptons have no impact on the flavor violating observable in the neutrino sector. For the matrix $ V_{PMNS} $, the standard parameterization form with parameters given by the neutrino
experiments \cite{epc59}. References \cite{12}\cite{219} have shown that, for $V_{Hl}=V_{PMNS}$, to make the $\mu\rightarrow e\gamma$ and $\mu^{-}\rightarrow e^{-}e^{+}e^{-}$ decay rates consistent with the present experimental upper bounds, the spectrum of the T-odd leptons (mirror leptons) must be quasi-degenerate.  So we will fix the mirror lepton masses ${M}_{Le}={M}_{L\mu}={M}_{L\tau}={M}_{L}$, and take the symmetry breaking scale $ f $ and the mirror lepton mass ${M}_{L}$ as free parameters. Furthermore, in order to make our numerical results more realistic,  we will apply the cut on the transverse momentum for radiated photon as $ P_{T}^{\gamma}> P_{T,cut}^{\gamma}$ with $P_{T,cut}^{\gamma}= 15 GeV $.
\begin{figure}[htbp]
\begin{center}
\includegraphics [scale=0.7] {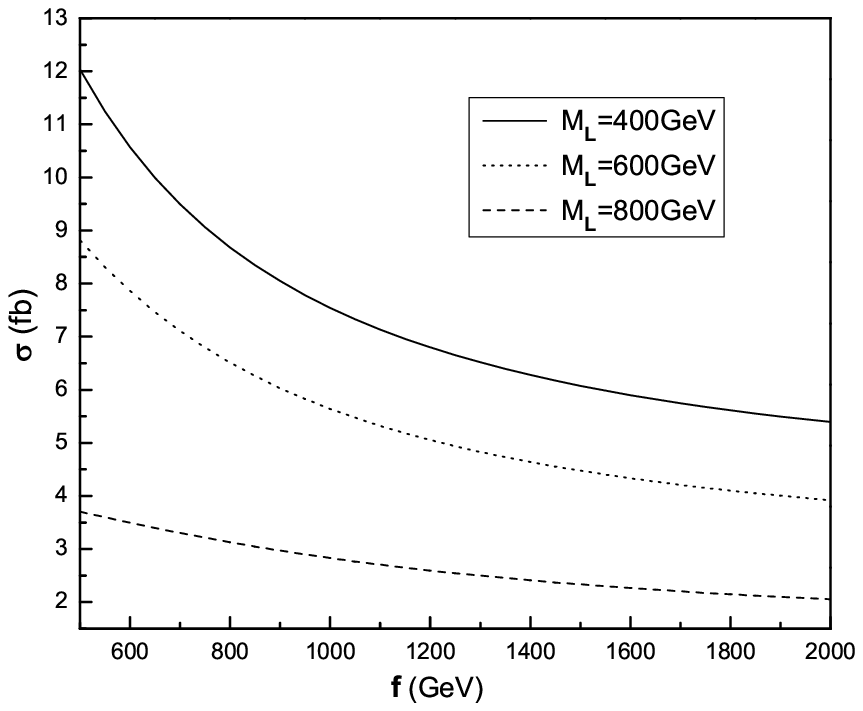}
\includegraphics [scale=0.7] {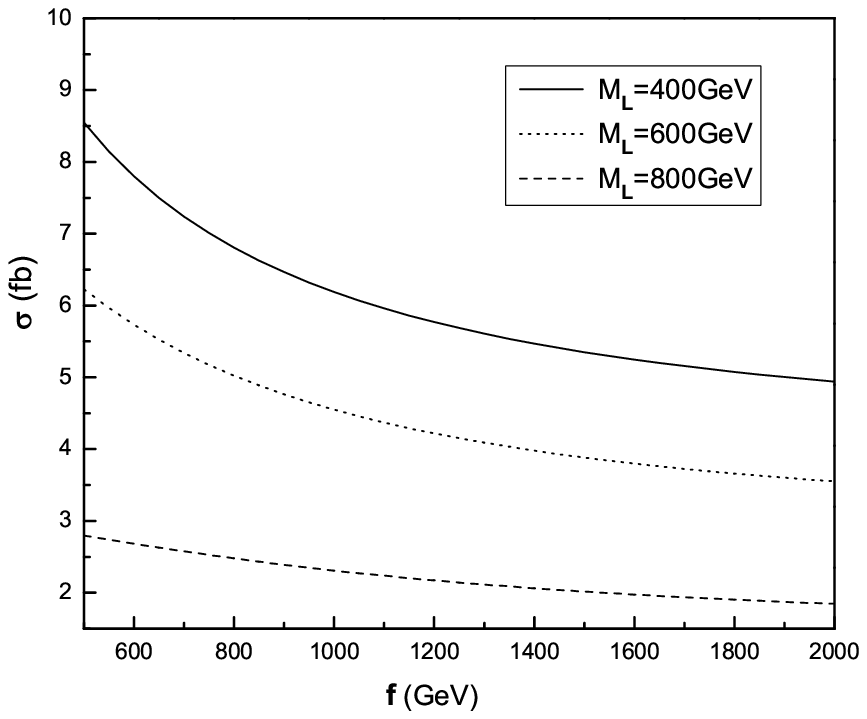}
\vspace{-0.2cm}(a)~~~~~~~~~~~~~~~~~~~~~~~~~~~~~~~~~~~~~~~~~~~~~~~~~~~~~~~~~~~~~~(b)
\vspace{-0.2cm}\caption{The production cross sections (a) $\sigma(\gamma\overline{L}_{e}{L}_{\mu})$ and  (b) $\sigma(\gamma\overline{L}_{\mu}{L}_{\mu})$ as  function of the scale \hspace*{1.6cm}parameter $f$  for $\sqrt{s}=2TeV$ and three values of the T-odd lepton mass ${M}_{L}$. }
\end{center}
\end{figure}

From Ref.\cite{epc59} we can find that the values of the matrix elements {$(V_{PMNS})_{e\tau}$} and {$(V_{PMNS})_{\tau e}$} are smaller than those of {$(V_{PMNS})_{e\mu}$} and {$(V_{PMNS})_{\mu e}$}, respectively. Therefore, it can be speculated that the production cross sections $\sigma(\gamma\overline{L}_{\tau}{L}_{\tau})$ and $\sigma(\gamma\overline{L}_{e}{L}_{\tau})$
 [ or $\sigma(\gamma\overline{L}_{\mu}{L}_{\tau})$] are smaller than $\sigma(\gamma\overline{L}_{\mu}{L}_{\mu})$ and $\sigma(\gamma\overline{L}_{e}{L}_{\mu})$, respectively. So we only give the cross sections $\sigma(\gamma\overline{L}_{\mu}{L}_{\mu})$ and $\sigma(\gamma\overline{L}_{e}{L}_{\mu})$ in the following calculation.
 The PMNS matrix $ V_{PMNS} $ have been constructed in Ref.\cite{epc59} based on PDG parametrization and the available data from oscillation experiments. To simply the numerical results, we take $(V_{PMNS})_{ee}=0.82$, $(V_{PMNS})_{\mu e}=0.50$ and $(V_{PMNS})_{e\mu}=0.55$.
\begin{figure}[htbp]
\begin{center}
\includegraphics [scale=0.7] {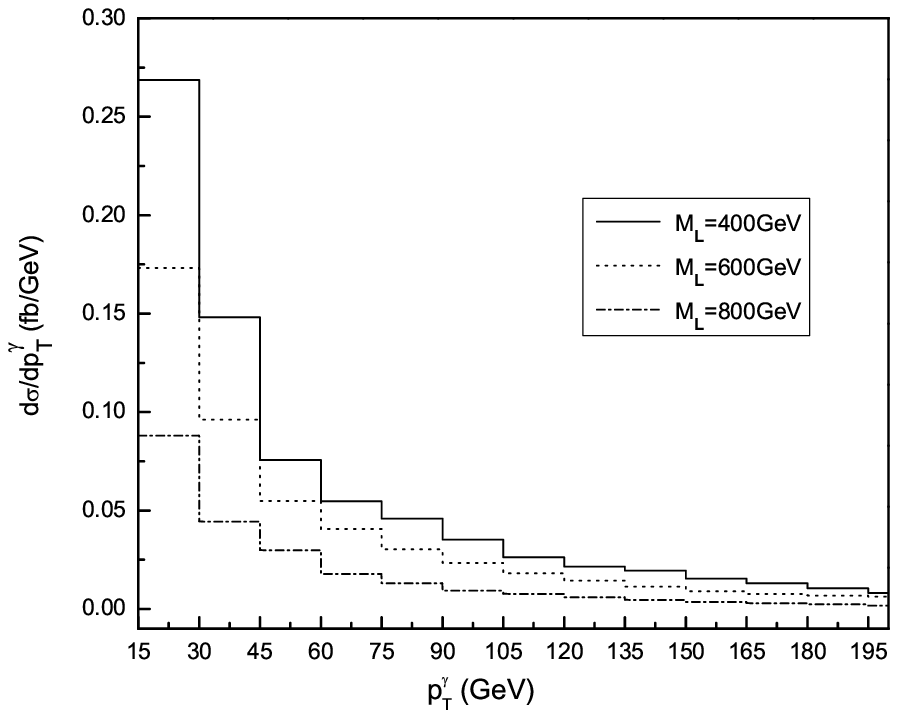}
\includegraphics [scale=0.7] {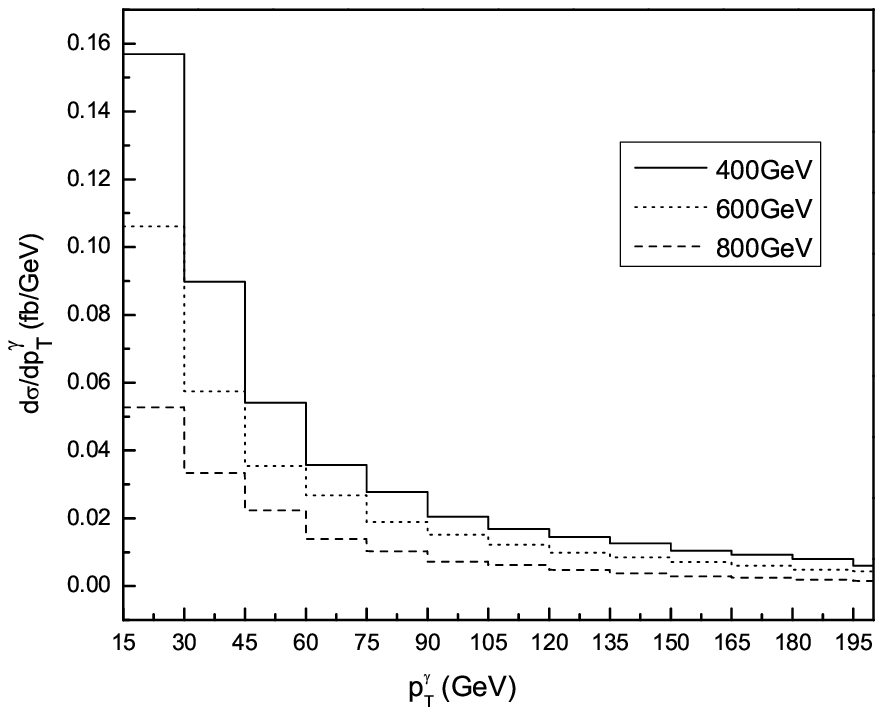}
(a)~~~~~~~~~~~~~~~~~~~~~~~~~~~~~~~~~~~~~~~~~~~~~~~~~~~~~~~~~~~~~~(b)
\vspace{-0.5cm}
\caption{The distributions of the transverse momentum of $\gamma$ photon $ P_{T}^{\gamma}$ for the $e^+e^-\rightarrow \gamma\overline{L}_{e}{L}_{\mu}$ \hspace*{1.6cm}process with (a)$f=0.5TeV$, (b)$f=1TeV$ for $\sqrt{s}=2TeV$  and three values of  \hspace*{1.6cm}the T-odd lepton mass ${M}_{L}$.}
\end{center}
\end{figure}

 In Fig.2, we plot the cross sections $\sigma(\gamma\overline{L}_{\mu}{L}_{\mu})$ and $\sigma(\gamma\overline{L}_{e}{L}_{\mu})$ as  function of the symmetry breaking scale $f$ for three values of the mass parameter ${M}_{L}$. The plots show that their values decrease as $f$ increases, which are in the ranges of $8.55-4.93fb$ and $11.96-5.39fb$, respectively, for ${M}_{L}$= $400GeV $ and $500GeV$ $\leq f \leq$ $2000GeV$. If we assume that the future ILC experiment has a yearly integrated luminosity of 100 $fb^{-1}$, then  several hundreds up to thousands of $\gamma\overline{L}_{i}{L}_{j}$ events will be generated per year.
\begin{figure}[htbp]
\begin{center}
\includegraphics [scale=0.7] {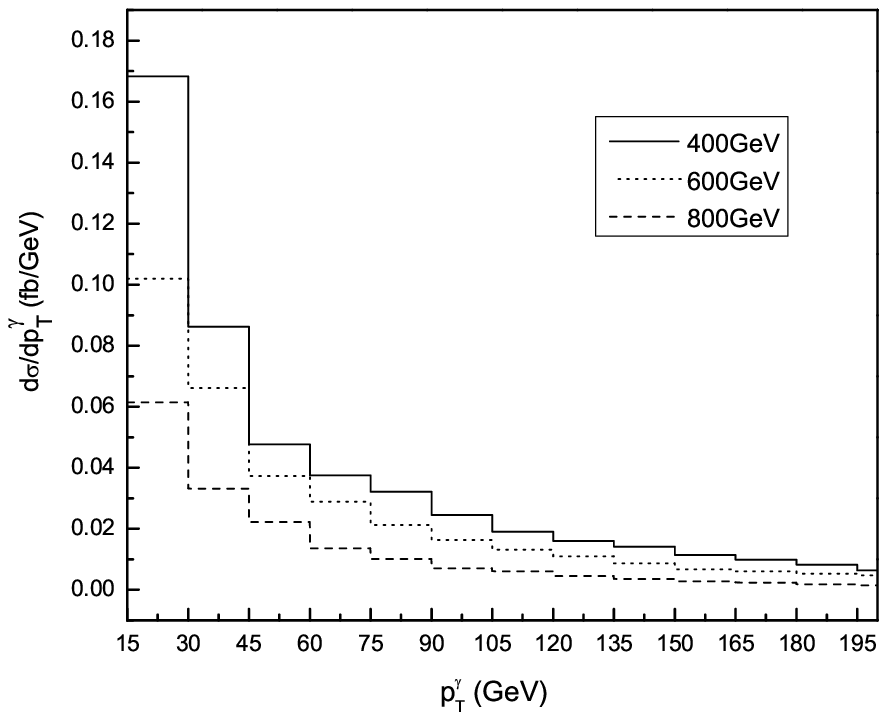}
\includegraphics [scale=0.7] {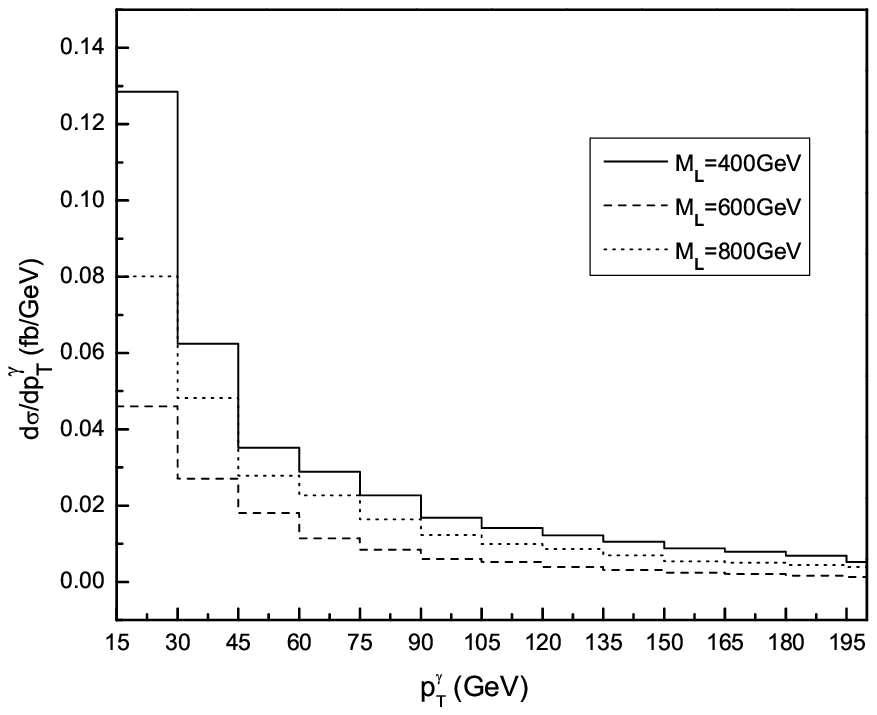}
(a)~~~~~~~~~~~~~~~~~~~~~~~~~~~~~~~~~~~~~~~~~~~~~~~~~~~~~~~~~~~~~~(b)
\caption{The distributions of the transverse momentum of $\gamma$ photon $ P_{T}^{\gamma}$ for the $e^+e^-\rightarrow \gamma\overline{L}_{\mu}{L}_{\mu}$ \hspace*{1.6cm}process with  (a)$f=0.5TeV$, (b)$f=1TeV$  for $\sqrt{s}=2TeV$, and three values of  \hspace*{1.6cm}the T-odd lepton mass ${M}_{L}$.}
\end{center}
\end{figure}

In Fig.3 and Fig.4, we plot the distribution of transverse momentum of the final state $\gamma$ photon for the processes $e^+e^-\rightarrow \gamma\overline{L}_{\mu}{L}_{\mu}$  and $e^+e^-\rightarrow \gamma\overline{L}_{e}{L}_{\mu}$ respectively, for $f = 500 GeV$ and $1000 GeV$ and three values of the mass parameter ${M}_{L}$. These figures illuminate that the symmetry breaking scale $f$ and mass parameter ${M}_{L}$ can significantly affect the values of differential cross sections ${d\sigma(\gamma\overline{L}_{\mu}{L}_{\mu})}/{dP_{T}^{\gamma}}$ and ${d\sigma(\gamma\overline{L}_{e}{L}_{\mu})}/{dP_{T}^{\gamma}}$. Their values increase quickly as $ P_{T}^{\gamma}$ decrease and most of the photons in the events of the processes $e^+e^-\rightarrow \gamma\overline{L}_{i}{L}_{j}$ are produced in the low transverse momentum range at the ILC.

To see whether the T-odd lepton ${L}_{i}$ can be observed at the ILC via the process $e^+e^-\rightarrow \gamma\overline{L}_{i}{L}_{j}$, we consider the possible decay modes of the T-odd lepton ${L}_{i}$. From Eqs.(1) and (2), we can see that, for  the Yukawa coupling constant $\kappa_i <0.46$, the T-odd lepton ${L}_{i}$ mainly decays to ${B}_{H}{l}_{i}$ (${l}_{i}=e,\mu$ or $\tau$) , while for $\kappa_i >0.46$, the T-odd leptons became heavier than the gauge bosons $W_{H}$ and $Z_{H}$ and other modes start opening up: ${W}_{H}{l}_{i}$ and ${Z}_{H}{l}_{i}$ \cite{71}. Furthermore, the mixing matrix $V_{Hl}$ allows  the FC decay  ${L}_{i}\rightarrow{B}_{H}{l}_{j}$ with $i$ different from $j$. The partial decay
width can be written in an unified manner as :
\begin {equation}
\Gamma(L\rightarrow lV_{H})=\frac{M^{3}_{L}g^{2}_{L}}{96\pi M_{V_{H}}}\{x^{2}(1-2x^{2}+y^{2})+{(1-y^{2}})^{2}\}\lambda^{\frac{1}{2}}(1,x^{2},y^{2}) ,
\end {equation}
with $x = M_{V_{H}}/M_{L} $, $y = M_{l}/M_{L}$, and  $\lambda(x, y, z) = x^{2}+ y^{2}+ z^{2}- 2xy - 2xz - 2yz$,
in which $M_{V_{H}}$ is the mass of the T-odd gauge boson. $g_{L}$ represents the coupling constant of the T-odd lepton ${L}$ to the T-odd gauge boson $V_{H}$ and the ordinary lepton $l$. For $M_{L}<M_{Z_{H}}\simeq M_{W_{H}}$,  the possible decay channels of the T-odd lepton ${L}_{i}$ are  ${L}_{i}\rightarrow {l}_{j}{B}_{H}$ (${l}_{j}=e,\mu$ or $\tau$).
For $f =1000 GeV$ and three values of the T-odd lepton mass ${M}_{L}=400GeV$, $600GeV$, and $800GeV$, the values of the total width for the decay channels $L_{e}\rightarrow l_{i}B_{H}$ are $1.94GeV$, $6.87GeV$, and $16.42GeV$, respectively. Similarly, for the decay channel $L_{\mu}\rightarrow l_{i}B_{H}$, the values of the total width are $1.95GeV$, $6.91GeV$, and $16.52GeV$, respectively. Furthermore, for $f =1000 GeV$, the branching ratios are $Br({L}_{e}\rightarrow \tau{B}_{H})=3\%$, $Br({L}_{e}\rightarrow \mu{B}_{H})=30\%$ and $Br({L}_{e}\rightarrow e{B}_{H})=67\%$, and $Br({L}_{\mu}\rightarrow e{B}_{H})=24.7\%$, $Br({L}_{\mu}\rightarrow\mu{B}_{H})=26.7\%$ and $Br({L}_{\mu}\rightarrow\tau{B}_{H})=48.6\%$.

 The new gauge boson ${B}_{H}$ which is the lightest T-odd particle can be seen as an attractive dark matter candidate \cite{10}.
 The decay modes of the T-odd lepton ${L}_{i}$ are  ${L}_{i}\rightarrow {l}_{j}{B}_{H}$ (${l}_{j}=e,\mu$ or $\tau$).
 Then the possible signatures of the process $e^+e^-\rightarrow \gamma\overline{L}_{e}{L}_{\mu}$ are the lepton flavor conservation final states $\gamma\overline{e}{e}$+$\rlap/E_{T}$, $\gamma\overline{\mu}{\mu}$+$\rlap/E_{T}$ and $\gamma\overline{\tau}{\tau}$+$\rlap/E_{T}$,  the lepton flavor violation final states  $\gamma\overline{e}{\tau}$+$\rlap/E_{T}$, $\gamma\overline{e}{\mu}$+$\rlap/E_{T}$ and $\gamma\overline{\mu}{\tau}$+$\rlap/E_{T}$. For the lepton flavor conservation signals, the intrinsic SM backgrounds mainly come from the processes $e^+e^-\rightarrow\gamma W^+W^-$ \cite{990331} with the $SM$ gauge bosons $W^{\pm}$  decay leptonically, $W^{\pm}\rightarrow l\nu$, and the process $e^+e^-\rightarrow\gamma ZZ$ \cite{990331} for one gauge boson $Z$ decaying to $\overline{l}l$ and another decaying to $\nu \overline{\nu}$. While for the lepton flavor violation signals, the main SM backgrounds  come from the process $e^+e^-\rightarrow\gamma W^+W^-$ for all of the SM gauge bosons $W^{\pm}$ leptonic decay.
For the T-odd leptons  $L_{e}$ and ${L}_{\mu}$, the decay channels with the largest branching rations are ${L}_{e}\rightarrow e{B}_{H}$ and ${L}_{\mu}\rightarrow\tau{B}_{H}$, respectively, so we will focus on these two decay modes in our following discussions.
The leading SM backgrounds of the largest production rate signals $\gamma\overline{e}{\tau}+\rlap/E_{T}$ come from the SM process $e^+e^-\rightarrow\gamma W^+W^-\rightarrow\gamma\overline{e}\tau{\nu}_{e}\overline{\nu}_{\tau}$. Using the results of the branching ratios $Br(W^{+}\rightarrow \overline{e}\nu_{e})$ and $Br(W^{-}\rightarrow \tau\overline{\nu}_{\tau})$ in \cite{16}, we recalculate  the cross section of the SM process $e^+e^-\rightarrow\gamma W^+W^-\rightarrow\gamma\overline{e}\tau{\nu}_{e}\overline{\nu}_{\tau}$, which is about $0.346fb$ at the ILC experiment with $\sqrt{s}=2TeV$. However, the cross section of the signal  $\gamma\overline{e}{\tau}$+$\rlap/E_{T}$ are large than $1.16fb$ in most range of the parameter space of the LHT model. Then the production rate of the intrinsic SM backgrounds is smaller than that generated by the process $e^+e^-\rightarrow \gamma\overline{L}_{e}{L}_{\mu}$.  So, the distinct signal ${\gamma\overline e\tau}+\rlap/E_{T}$ should be easily separated from the SM backgrounds. The possible signatures of the process $e^+e^-\rightarrow \gamma\overline{L}_{\mu}{L}_{\mu}$ are  same as those of the process $e^+e^-\rightarrow \gamma\overline{L}_{e}{L}_{\mu}$, but, the production rates of the signatures are different for these two processes.
For the process $e^+e^-\rightarrow \gamma\overline{L}_{\mu}{L}_{\mu}$, the signal with largest production rate  is ${\gamma\overline\tau}{\tau}$ plus large missing energy $\rlap/E_{T}$, ${\gamma\overline\tau}{\tau}$+$\rlap/E_{T}$. Its intrinsic SM backgrounds mainly come from the processes $e^+e^-\rightarrow\gamma W^+W^-\rightarrow\gamma\tau\overline{\tau}{\nu}_{\tau}\overline{\nu}_{\tau}$ and $e^+e^-\rightarrow\gamma ZZ\rightarrow\gamma\tau\overline{\tau}\nu\overline{\nu}$ with ${\nu} ={\nu}_{e},{\nu}_{\mu},{\nu}_{\tau}$. At the ILC experiment with $\sqrt{s}=2TeV$,  we recalculate their cross sections and find that their values are about $0.362fb$ and $0.0094fb$, respectively.  However, the cross section of the signal  $\gamma\overline{\tau}{\tau}$+$\rlap/E_{T}$ are larger than $0.682fb$ in most of the parameter space of the LHT model.
Thus, it may be possible to extract the signals from the backgrounds in the reasonable parameters space of the LHT model.

From the above discussions,  we can see that, considering the FC decay ${L}_{i}\rightarrow{B}_{H}{l}_{j}(i\neq j)$, both of the processes $e^+e^-\rightarrow \gamma\overline{L}_{e}{L}_{\mu}$ and $e^+e^-\rightarrow \gamma\overline{L}_{\mu}{L}_{\mu}$ can give rise to the signals $\gamma\overline{e}{\tau}$+$\rlap/E_{T}$ and ${\gamma\overline\tau}{\tau}$+$\rlap/E_{T}$. Thus, the FC decay ${L}_{i}\rightarrow{B}_{H}{l}_{j}(i\neq j)$  generates some interplay  between $\gamma\overline{L}_{e}{L}_{\mu}$ and $ \gamma\overline{L}_{\mu}{L}_{\mu}$ intermediate states. This interference effect enhances the observability of the signals and further strengthens our physical conclusions. It should be noted that this conclusion also apply other signals generated by he processes $e^+e^-\rightarrow \gamma\overline{L}_{e}{L}_{\mu}$ and $e^+e^-\rightarrow \gamma\overline{L}_{\mu}{L}_{\mu}$.

\section{THE PROCESS $e^+e^-\rightarrow Z\overline{L}_{i}{L}_{j}$ }
\noindent

  The T-odd leptons can also be produced in association with a gauge boson $Z$  at the ILC as shown in Fig.1. Based on the Feynman rules of the T-odd leptons written as above, the invariant production amplitude of the process $e^+(p_{1})e^-(p_{2})\rightarrow Z(p_5)\overline{L}_{i}(p_{4}){L}_{j}(p_{3})$  can be written as
\begin{eqnarray}
\mathcal
{M}_{2}&=&{M}^{\gamma Z}_{a}+{M}^{ZZ}_{a}+{M}^{\gamma Z}_{b}+{M}^{ZZ}_{b}
+{M}^{\gamma Z}_{c}+{M}^{ZZ}_{c}+{M}^{\gamma Z}_{d}+{M}^{ZZ}_{d}+{M}^{B_{H}Z}_{e}\nonumber\\&&
+{M}^{Z_{H}Z}_{e}+{M}^{B_{H}Z}_{f}+{M}^{Z_{H}Z}_{f}
+{M}^{B_{H}Z}_{g}+{M}^{Z_{H}Z}_{g}+{M}^{B_{H}Z}_{h}+{M}^{Z_{H}Z}_{h}
\end{eqnarray}
with
\begin{eqnarray}
 {M}^{\gamma Z}_{a}&=&\frac{-ie^3}{S_WC_W}(-\frac{1}{2}+S^2_W)G(p_1+p_2,0)G(p_4+p_5,{M}_{L})\bar{v}(p_{1})\gamma^{\mu}u(p_{2})\bar{u}(p_{3})
 \gamma_{\mu}\nonumber\\&&\times[-(\pslash_4+\pslash_5)+{M}_{L}]\rlap/\epsilon(p_5)v(p_{4}),
\end{eqnarray}
\begin{eqnarray}
 {M}^{ZZ}_{a}&=&\frac{-ie^3}{S^3_WC^3_W}(-\frac{1}{2}+S^2_W)^2G(p_1+p_2,{M}_{Z})G(p_4+p_5,{M}_{L})\bar{v}(p_{1})\gamma^{\mu}\nonumber\\&&
 \times[(-\frac{1}{2}+S^{2}_W)P_{L}+(S^{2}_W)P_{R}]
 u(p_{2})\bar{u}(p_{3})\gamma_{\mu}[-(\pslash_4+\pslash_5)+{M}_{L}]\rlap/\epsilon(p_5)v(p_{4}),
\end{eqnarray}
 \begin{eqnarray}
 {M}^{\gamma Z}_{b}&=&\frac{-ie^3}{S_WC_W}(-\frac{1}{2}+S^2_W)G(p_1+p_2,0)G(p_3+p_5,{M}_{L})
 \bar{v}(p_{1})\gamma^{\mu}u(p_{2})\bar{u}(p_{3})\rlap/\epsilon(p_5)\nonumber\\&&\times[\pslash_3+\pslash_5+{M}_{L}]\gamma_{\mu}v(p_{4}),
\end{eqnarray}
\begin{eqnarray}
 {M}^{ZZ}_{b}&=&\frac{-ie^3}{S^3_WC^3_W}(-\frac{1}{2}+S^2_W)^2G(p_1+p_2,{M}_{Z})G(p_3+p_5,{M}_{L})
 \bar{v}(p_{1})\gamma^{\mu}[(-\frac{1}{2}+S^{2}_W)P_{L}\nonumber\\&&
 +(S^{2}_W)P_{R}]u(p_{2})\bar{u}(p_{3})\rlap/\epsilon(p_5)[\pslash_3+\pslash_5+{M}_{L}]\gamma_{\mu}v(p_{4}),
\end{eqnarray}
\begin{eqnarray}
 {M}^{\gamma Z}_{c}&=&\frac{-ie^3}{S_WC_W}G(p_1-p_5,{M}_{e})G(p_3+p_4,0)\bar{v}(p_{1})\rlap/\epsilon(p_5)
 [(-\frac{1}{2}+S^{2}_W)P_{L}+(S^{2}_W)P_{R}]\nonumber\\&&\times [-(\pslash_1-\pslash_5)+{M}_{e}]\gamma^{\mu}u(p_{2})\bar{u}(p_{3})
 \gamma_{\mu}v(p_{4}),
\end{eqnarray}
\begin{eqnarray}
 {M}^{ZZ}_{c}&=&\frac{-ie^3}{S^3_WC^3_W}(-\frac{1}{2}+S^2_W)G(p_1-p_5,{M}_{e})G(p_3+p_4,{M}_{Z})\bar{v}(p_{1})\rlap/\epsilon(p_5)
 [(-\frac{1}{2}+S^{2}_W)P_{L}+\nonumber\\&&(S^{2}_W)P_{R}][-(\pslash_1-\pslash_5)+{M}_{e}]
 \gamma^{\mu}[(-\frac{1}{2}+S^{2}_W)P_{L}+(S^{2}_W)P_{R}] u(p_{2})\bar{u}(p_{3})\gamma_{\mu}v(p_{4}),
\end{eqnarray}
\begin{eqnarray}
 {M}^{\gamma Z}_{d}&=&\frac{-ie^3}{S_WC_W}G(p_2-p_5,{M}_{e})G(p_3+p_4,0)\bar{v}(p_{1})\gamma^{\mu} [(\pslash_2-\pslash_5)+{M}_{e}]\rlap/\epsilon(p_5)[(-\frac{1}{2}+S^{2}_W)P_{L} \nonumber\\&& +(S^{2}_W)P_{R}]
u(p_{2})\bar{u}(p_{3})\gamma_{\mu}v(p_{4}),
\end{eqnarray}
\begin{eqnarray}
 {M}^{ZZ}_{d}&=&\frac{-ie^3}{S^3_WC^3_W}G(p_2-p_5,{M}_{e})G(p_3+p_4,{M}_{Z})(-\frac{1}{2}+S^2_W)
 \bar{v}(p_{1})\gamma^{\mu}[(-\frac{1}{2}+S^{2}_W)P_{L}+S^{2}_WP_{R}]\nonumber\\&&
 [(\pslash_2-\pslash_5)+{M}_{e}]\rlap/\epsilon(p_5)
 [(-\frac{1}{2}+S^{2}_W)P_{L}+(S^{2}_W)P_{R}]
u(p_{2})\bar{u}(p_{3})\gamma_{\mu}v(p_{4}),
\end{eqnarray}
\begin{eqnarray}
 {M}^{{B}_{H}Z}_{e}&=&\frac{-ie^3}{S_WC^3_W}(-\frac{1}{2}+S^2_W)[\frac{1}{10}+\frac{5C^2_W}{8(5C^2_W-S^2_W)}\frac{\nu^2}{f^2}]^2(V_{Hl})_{ie}(V_{Hl})_{ej}
 \nonumber\\&&
 \times G(p_4+p_5,{M}_{L})G(p_2-p_3,{M}_{{B_{H}}})
 \bar{v}(p_{1})\gamma^{\mu}P_{L}[-(\pslash_4+\pslash_5)+{M}_{L}]\nonumber\\&&
 \times \rlap/\epsilon(p_5)v_(p_{4})\bar{u}(p_{3})\gamma_{\mu}P_{L}u(p_{2}),
\end{eqnarray}
\begin{eqnarray}
 {M}^{{Z}_{H}Z}_{e}&=&\frac{-ie^3}{S^3_WC_W}(-\frac{1}{2}+S^2_W)[-\frac{1}{2}+\frac{S^2_W}{8(5C^2_W-S^2_W)}\frac{\nu^2}{f^2}]^2(V_{Hl})_{ie}(V_{Hl})_{ej}
 \nonumber\\&&
 \times G(p_4+p_5,{M}_{L})G(p_2-p_3,{M}_{{Z_{H}}})
 \bar{v}(p_{1})\gamma^{\mu}P_{L}[-(\pslash_4+\pslash_5)+{M}_{L}]\nonumber\\&&
 \times \rlap/\epsilon(p_5)v(p_{4})\bar{u}(p_{3})\gamma_{\mu}P_{L}u(p_{2}),
\end{eqnarray}
\begin{eqnarray}
 {M}^{{B}_{H}Z}_{f}&=&\frac{-ie^3}{S_WC^3_W}(-\frac{1}{2}+S^2_W)[\frac{1}{10}+\frac{5C^2_W}{8(5C^2_W-S^2_W)}\frac{\nu^2}{f^2}]^2(V_{Hl})_{ie}(V_{Hl})_{ej}
 G(p_4-p_1,{M}_{{B}_{H}})\nonumber\\&&
 \times G(p_3+p_5,{M}_{L})
 \bar{v}(p_{1})\gamma^{\mu}P_{L}v(p_{4})\bar{u}(p_{3})\rlap/\epsilon(p_5)[\pslash_3+\pslash_5+{M}_{L}]\gamma_{\mu}P_{L}u(p_{2}),
\end{eqnarray}
\begin{eqnarray}
 {M}^{{Z}_{H}Z}_{f}&=&\frac{-ie^3}{S^3_WC_W}(-\frac{1}{2}+S^2_W)[-\frac{1}{2}+\frac{S^2_W}{8(5C^2_W-S^2_W)}\frac{\nu^2}{f^2}]^2(V_{Hl})_{ie}(V_{Hl})_{ej}
 G(p_4-p_1,{M}_{{Z}_{H}})\nonumber\\&&
 \times G(p_3+p_5,{M}_{L})
 \bar{v}(p_{1})\gamma^{\mu}P_{L}v(p_{4})\bar{u}(p_{3})\rlap/\epsilon(p_5)[\pslash_3+\pslash_5+{M}_{L}]\gamma_{\mu}P_{L}u(p_{2}),
\end{eqnarray}
\begin{eqnarray}
 {M}^{{B}_{H}Z}_{g}&=&\frac{-ie^3}{S_WC^3_W}[\frac{1}{10}+\frac{5C^2_W}{8(5C^2_W-S^2_W)}\frac{\nu^2}{f^2}]^2(V_{Hl})_{ie}(V_{Hl})_{ej}
 G(p_5-p_1,{M}_{e})G(p_2-p_3,{M}_{{B_{H}}})
 \nonumber\\&&\times
 \bar{v}(p_{1})\rlap/\epsilon(p_5)(-\frac{1}{2}P_{L}+S^2_W)[\pslash_5-\pslash_1+{M}_{e}]
 \gamma^{\mu}P_{L}v_(p_{4})\bar{u}(p_{3})\gamma_{\mu}P_{L}u(p_{2}),
\end{eqnarray}
\begin{eqnarray}
 {M}^{{Z}_{H}Z}_{g}&=&\frac{-ie^3}{S_W^3C_W}[-\frac{1}{2}+\frac{S^2_W}{8(5C^2_W-S^2_W)}\frac{\nu^2}{f^2}]^2(V_{Hl})_{ie}(V_{Hl})_{ej}
 G(p_5-p_1,{M}_{e})G(p_2-p_3,{M}_{{Z_{H}}})
 \nonumber\\&&\times
 \bar{v}(p_{1})\rlap/\epsilon(p_5)(-\frac{1}{2}P_{L}+S^2_W)[\pslash_5-\pslash_1+{M}_{e}]
 \gamma^{\mu}P_{L}v_(p_{4})\bar{u}(p_{3})\gamma_{\mu}P_{L}u(p_{2}),
\end{eqnarray}
\begin{eqnarray}
 {M}^{{B}_{H}Z}_{h}&=&\frac{-ie^3}{S_WC^3_W}[\frac{1}{10}+\frac{5C^2_W}{8(5C^2_W-S^2_W)}\frac{\nu^2}{f^2}]^2(V_{Hl})_{ie}(V_{Hl})_{ej}
 G(p_4-p_1,{M}_{{B}_{H}}) G(p_2-p_5,{M}_{e})\nonumber\\&& \times
 \bar{v}(p_{1})\gamma^{\mu}P_{L}v(p_{4})\bar{u}(p_{3})\gamma_{\mu}P_{L}[\pslash_2-\pslash_5+{M}_{e}]
 \rlap/\epsilon(p_5)(-\frac{1}{2}P_{L}+S^2_W)u(p_{2}),
\end{eqnarray}
\begin{eqnarray}
 {M}^{{Z}_{H}Z}_{h}&=&\frac{-ie^3}{S_W^3C_W}[-\frac{1}{2}+\frac{S^2_W}{8(5C^2_W-S^2_W)}\frac{\nu^2}{f^2}]^2(V_{Hl})_{ie}(V_{Hl})_{ej}
 G(p_4-p_1,{M}_{{Z}_{H}}) G(p_2-p_5,{M}_{e})\nonumber\\&& \times
 \bar{v}(p_{1})\gamma^{\mu}P_{L}v(p_{4})\bar{u}(p_{3})\gamma_{\mu}P_{L}[\pslash_2-\pslash_5+{M}_{e}]
 \rlap/\epsilon(p_5)(-\frac{1}{2}P_{L}+S^2_W)u(p_{2}).
\end{eqnarray}
  \begin{figure}[htbp]
\begin{center}
\includegraphics [scale=0.7] {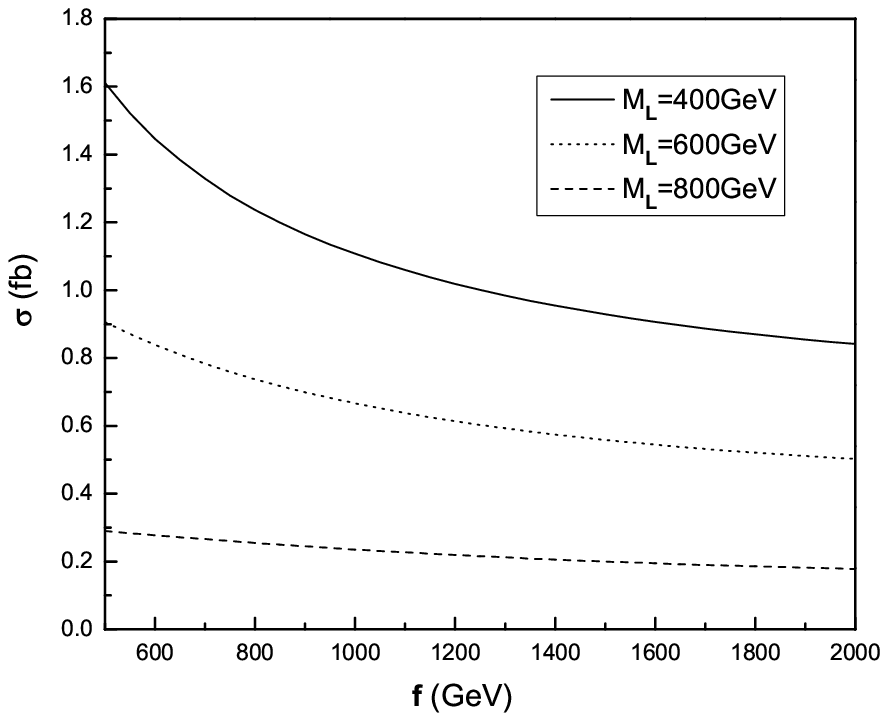}
\includegraphics [scale=0.7] {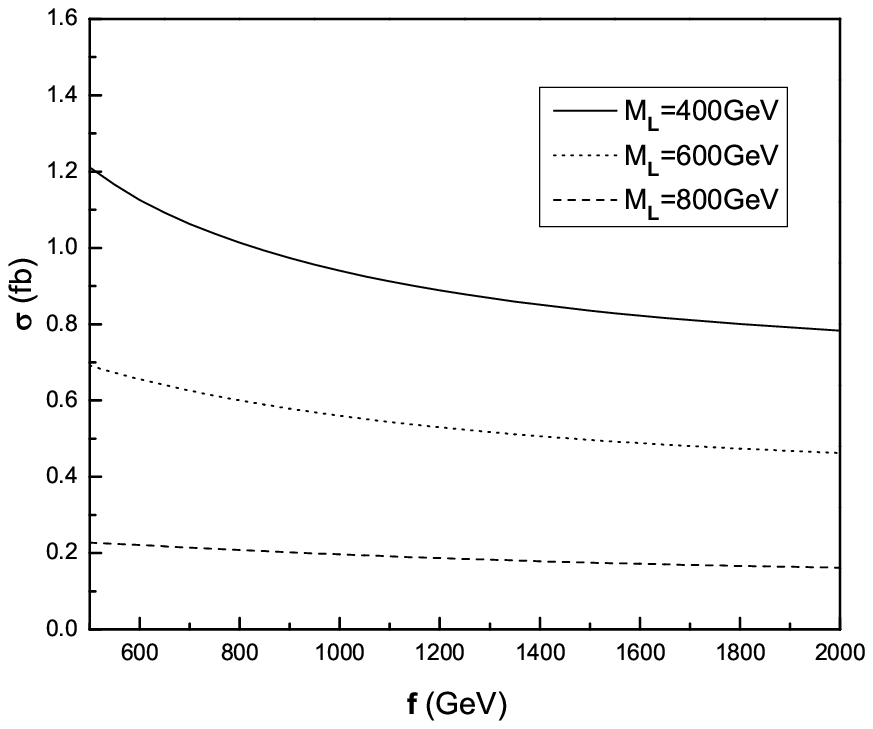}
(a)~~~~~~~~~~~~~~~~~~~~~~~~~~~~~~~~~~~~~~~~~~~~~~~~~~~~~~~~~~~~~~(b)
\caption{The production cross sections (a) $\sigma(Z\overline{L}_{e}{L}_{\mu})$ and (b) $\sigma(Z\overline{L}_{\mu}{L}_{\mu})$ as function of the scale \hspace*{1.6cm}parameter $f$  for $\sqrt{s}=2TeV$ and three values of the T-odd lepton mass ${M}_{L}$.}
\end{center}
\end{figure}

Similar with above, we can give the numerical results about the process $e^+e^-\rightarrow Z\overline{L}_{i}{L}_{j}$, which are summarized in Fig.5. We plot the cross sections of the processes $e^+e^-\rightarrow Z\overline{L}_{e}{L}_{\mu}$ and $e^+e^-\rightarrow Z\overline{L}_{\mu}{L}_{\mu}$  as  functions of the symmetry breaking scale $f$ for three values of  the mirror lepton mass ${M}_{L}$ in Fig.5.
One can see  that the cross sections $\sigma(Z\overline{L}_{e}{L}_{\mu})$  and $\sigma(Z\overline{L}_{\mu}{L}_{\mu})$ fall sharply as $f$ increases for fixed T-odd lepton mass ${M}_{L}$ and their values are also sensitive to the mass of the T-odd leptons, which is similar with those of the processes $e^+e^-\rightarrow \gamma\overline{L}_{e}{L}_{\mu}$ and $e^+e^-\rightarrow \gamma\overline{L}_{\mu}{L}_{\mu}$. This is because the phase space is depressed strongly by large final states and the coupling are related to the factor $\nu^{2}/f^2$.  The cross sections $\sigma(Z\overline{L}_{e}{L}_{\mu})$ and $\sigma(Z\overline{L}_{\mu}{L}_{\mu})$ are smaller than those of the processes $e^+e^-\rightarrow \gamma\overline{L}_{e}{L}_{\mu}$ and $e^+e^-\rightarrow \gamma\overline{L}_{\mu}{L}_{\mu}$, respectively. For $f = 500-2000GeV$ and ${M}_{L} = 400-800GeV$, the value of cross section $\sigma(Z\overline{L}_{e}{L}_{\mu})$[$\sigma(Z\overline{L}_{\mu}{L}_{\mu})$] is in the range of $1.62-0.187fb$[$1.21-0.171fb$].
\begin{figure}[htbp]
\begin{center}
\includegraphics [scale=0.7] {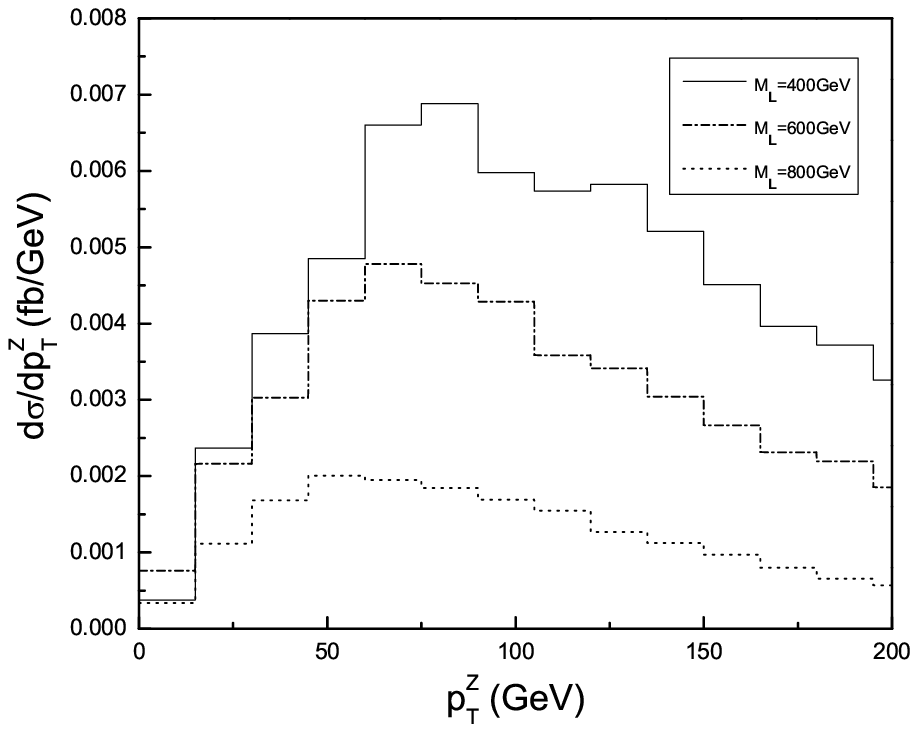}
\includegraphics [scale=0.7] {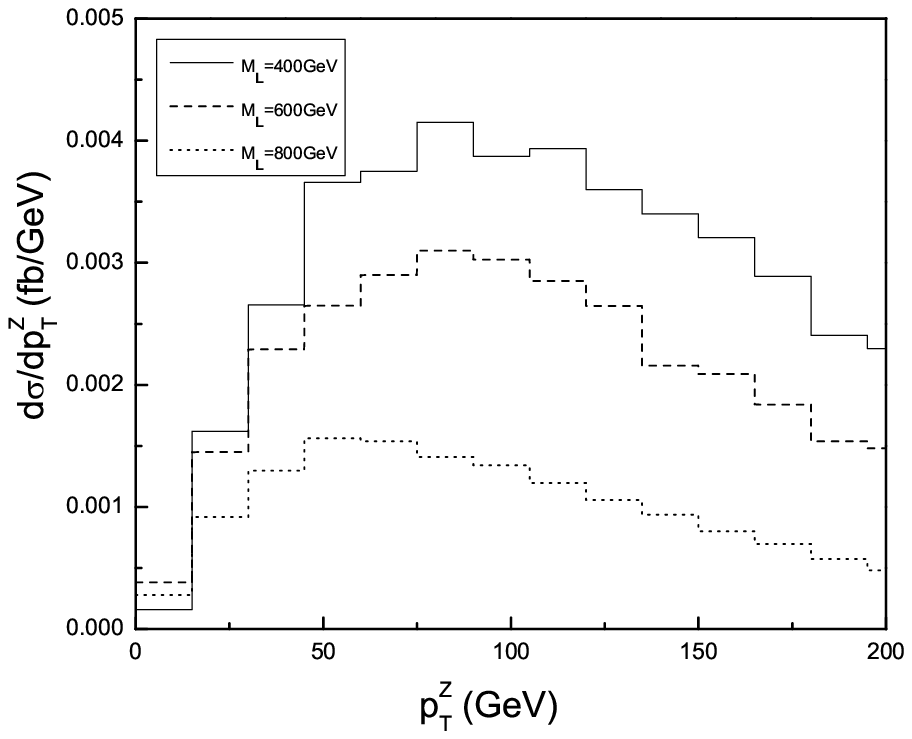}
(a)~~~~~~~~~~~~~~~~~~~~~~~~~~~~~~~~~~~~~~~~~~~~~~~~~~~~~~~~~~~~~~(b)
\caption{The distributions of the transverse momentum of $Z$ boson $ P_{T}^Z$ for the $e^+e^-\rightarrow Z\overline{L}_{e}{L}_{\mu}$  \hspace*{1.6cm}process with $f=0.5TeV$ (a) and $f=1TeV$ (b) for $\sqrt{s}=2TeV$, and three values of  \hspace*{1.6cm}the T-odd lepton mass ${M}_{L}$.}
\end{center}
\end{figure}
 The distributions of the transverse momentum of the gauge boson $Z$  are depicted in Fig.6 and Fig.7 corresponding to $f =0.5 TeV$ and $1 TeV$, respectively. From these two figures we can see that there exist peaks at different conditions, there are significant regions of $ P_{T}^{Z}$  with differential values of the symmetry breaking scale $f$ and the mirror lepton mass  ${M}_{L}$. The larger values of the symmetry breaking scale $f$ and the mirror lepton mass ${M}_{L}$ can significantly suppress the cross sections ${d\sigma(Z\overline{L}_{\mu}{L}_{\mu})}/{dP_{T}^{Z}}$ and ${d\sigma(Z\overline{L}_{e}{L}_{\mu})}/{dP_{T}^{Z}}$. Obviously, for $f \geq 1TeV$ and ${M}_{L} \geq 800GeV$, the values of cross sections are quite small.
\begin{figure}[htbp]
\begin{center}
\includegraphics [scale=0.68] {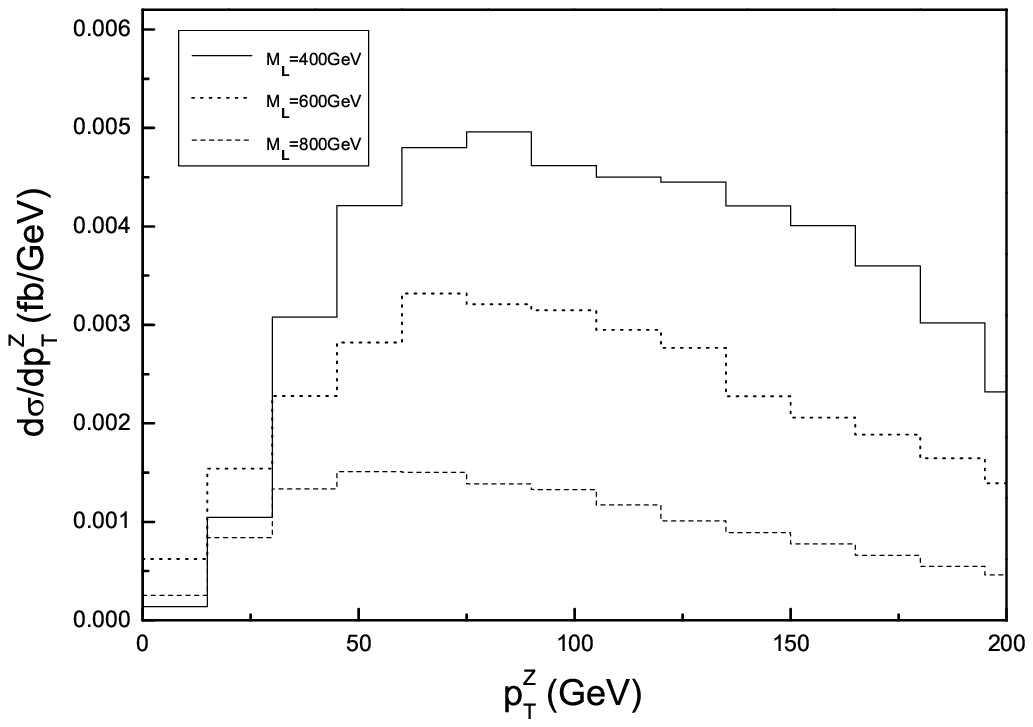}
\includegraphics [scale=0.68] {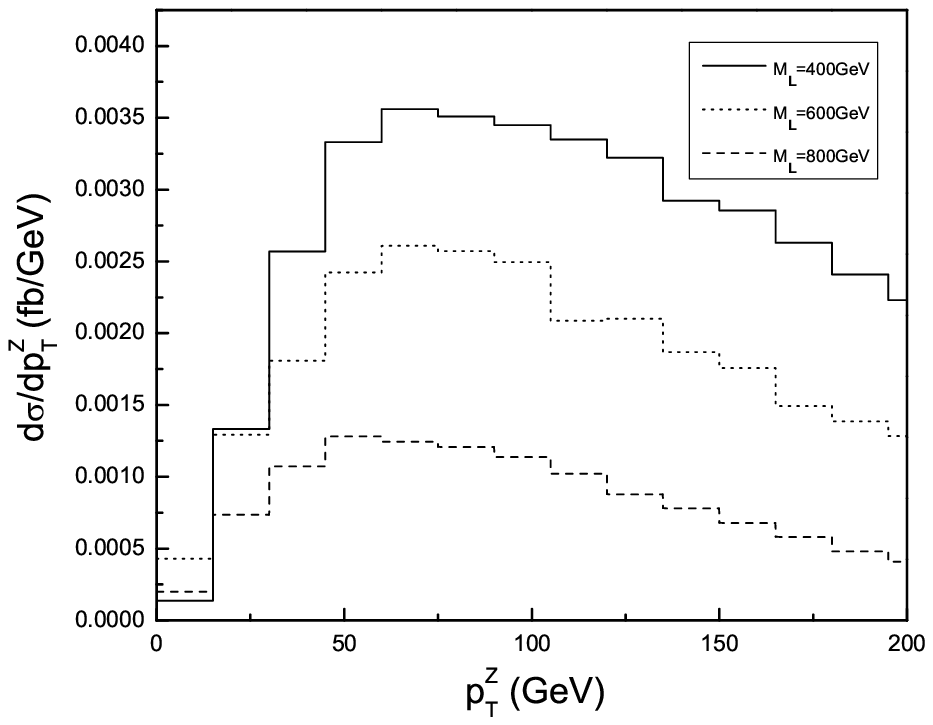}
(a)~~~~~~~~~~~~~~~~~~~~~~~~~~~~~~~~~~~~~~~~~~~~~~~~~~~~~~~~~~~~~~(b)
\caption{The distributions of the transverse momentum of $Z$ boson $ P_{T}^Z$ for the $e^+e^-\rightarrow Z\overline{L}_{\mu}{L}_{\mu}$ \hspace*{1.6cm}process with $f=0.5TeV$ (a) and $f=1TeV$ (b) for $\sqrt{s}=2TeV$, and three values of  \hspace*{1.6cm}the T-odd lepton mass ${M}_{L}$.}
\end{center}
\end{figure}

If we assume the final state $Z$ decaying to $\overline{l}l$, the possible signatures of the process $e^+e^-\rightarrow Z\overline{L}_{e}{L}_{\mu}$ are the lepton flavor conservation final states $\overline{l}l\overline{e}{e}$+$\rlap/E_{T}$, $\overline{l}l\overline{\mu}{\mu}$+$\rlap/E_{T}$ and $\overline{l}l\overline{\tau}{\tau}$+$\rlap/E_{T}$, and the lepton flavor violation final states  $\overline{l}l\overline{e}{\tau}$+$\rlap/E_{T}$, $\overline{l}l\overline{e}{\mu}$+$\rlap/E_{T}$ and $\overline{l}l\overline{\mu}{\tau}$+$\rlap/E_{T}$.
For the lepton flavor conservation signals, the intrinsic SM backgrounds mainly come from the processes $e^+e^-\rightarrow ZW^+W^-$ \cite{smz} with the gauge bosons $W^{\pm}$  decay leptonically, $W^{\pm}\rightarrow l\nu$, and the process $e^+e^-\rightarrow ZZW^+W^-$ for  one gauge boson $Z$ decaying to $\overline{l}l$ and other decaying to $\nu \overline{\nu}$.

While for the lepton flavor violation signals, the main SM backgrounds  come from the processes $e^+e^-\rightarrow ZW^+W^-$ with  the gauge boson $Z$ decaying to $\overline{l}l$ and the  gauge bosons $W^{\pm}$ leptonic decay, and the process $e^+e^-\rightarrow ZZW^+W^-$  for all of the  gauge bosons $W^{\pm}$ leptonic decay, and one gauge boson $Z$ decaying to $\overline{l}l$ another decaying to $\nu \overline{\nu}$.
From discussions given by section II, it is obviously that,  the largest production rate signal of the process $e^+e^-\rightarrow Z\overline{L}_{e}{L}_{\mu}$  is  $ \overline{l}l\overline{e}{\tau}$+$\rlap/E_{T}$. The leading SM backgrounds of this kind of signal  come from the SM processes $e^+e^-\rightarrow ZW^+W^-\rightarrow \overline{l}l\overline{e}\tau{\nu}_{e}\overline{\nu}_{\tau}$, and $e^+e^-\rightarrow ZZW^+W^-\rightarrow \overline{l}l\overline{e}\tau{\nu}_{e}\overline{\nu}_{\tau}{\nu}\overline{\nu}$ $({\nu} ={\nu}_{e},{\nu}_{\mu},{\nu}_{\tau})$.
Our numerical results indicate that, in wide range of the parameter space of the LHT model, the value of the statistical significance $S/\sqrt[]{B}$ is larger than $5$. In our numerical estimation, we have taken the integrated luminosity $\pounds= 100fb^{-1}$ and $\sqrt{S}=2TeV$. Thus, it may be possible to extract the signals from the backgrounds in the reasonable parameter space of the LHT model.
The possible signatures of the process $e^+e^-\rightarrow Z\overline{L}_{\mu}{L}_{\mu}$ are  same as that of the process $e^+e^-\rightarrow Z\overline{L}_{e}{L}_{\mu}$, but, the production rates are different from those generated by the process $e^+e^-\rightarrow Z\overline{L}_{e}{L}_{\mu}$.
For the process $e^+e^-\rightarrow Z\overline{L}_{\mu}{L}_{\mu}$, the signal with largest production rate  is the same-flavor opposite-sign pair leptons  ${\overline{l}{l}\overline\tau}{\tau}$ plus large missing energy $\rlap/E_{T}$; i.e. ${\overline{l}{l}\overline\tau}{\tau}$+$\rlap/E_{T}$. The SM backgrounds mainly come from the $e^+e^-\rightarrow ZW^+W^-\rightarrow\overline{l}{l}\tau\overline{\tau}{\nu}_{\tau}\overline{\nu}_{\tau}$ and $e^+e^-\rightarrow ZZW^+W^-\rightarrow\overline{l}{l}{\tau}\overline{\tau}{\nu}_{\tau}\overline{\nu}_{\tau}\nu\overline{\nu}$ $({\nu} ={\nu}_{e},{\nu}_{\mu},{\nu}_{\tau})$. Our numerical results indicate that,  the value of the statistical significance $S/\sqrt[]{B}$ is larger than $3$ in wide range of the parameter space of the LHT model. Thus, as long as the T-odd leptons are not too heavy, its possible signals might be detected via the processes $e^+e^-\rightarrow Z\overline{L}_{i}{L}_{j}$ in the future ILC experiments.

Certainly, the SM backgrounds must be further studied, detailed confirmation of the observability of the signals generated by the process $e^+e^-\rightarrow V\overline{L}_{i}{L}_{j}$ would require more Monte Carlo simulations of the signals and backgrounds, which is beyond the scope of this paper.

\section{Conclusions} \noindent

The LHT model is one of the attractive little Higgs models, which not only is consistent with EW precision tests but also provides a possible dark matter candidate. The heavy T-odd  fermions (mirror leptons and mirror quarks) are introduced to implement T-parity in the fermion sector of the model. These new heavy fermions might produce the observability signatures in future high energy collider experiments.

In this paper we consider pair production of the T-odd leptons in association with a gauge boson $V(=\gamma$ or $Z)$  in the future ILC experiments. The production cross sections of these processes and their  distributions of the transverse momentum are calculated. Our numerical results show that the cross section of the process $e^+e^-\rightarrow \gamma\overline{L}_{i}{L}_{j}$ is larger than that of the process $e^+e^-\rightarrow Z \overline{L}_{i}{L}_{j}$. For $\sqrt{s}=2TeV$, ${M}_{L} = 400 - 800GeV$ and $f = 500 - 2000GeV$, the values of the cross sections $\sigma(\gamma\overline{L}_{e}{L}_{\mu})$ and $\sigma(\gamma\overline{L}_{\mu}{L}_{\mu})$ are in the ranges of $11.96- 2.16fb$ and $8.55- 1.89fb$, while those for $\sigma(Z\overline{L}_{e}{L}_{\mu})$ and $\sigma(Z\overline{L}_{\mu}{L}_{\mu})$ are in the ranges of $1.62 - 0.187fb$ and $1.21 - 0.171fb$. We further analyze their possible signals and the corresponding SM backgrounds, also calculate the value of the statistical significance $S/\sqrt[]{B}$ for some processes.
 We find that, as long as the T-odd leptons are not too heavy, they can be copiously produced via the processes
$e^+e^-\rightarrow V\overline{L}_{i}{L}_{j}$, and their signatures might be observed in the future ILC experiments. Thus, we expect that these production processes can be used to detect the T-odd leptons predicted by the LHT model in the future ILC experiments.
\vspace{4mm}
\\
\vspace{4mm} \textbf{Acknowledgments}\\

 This work was supported in part by the National Natural Science Foundation of
China under Grants No.10975067, the Specialized Research Fund for
the Doctoral Program of Higher Education (SRFDP) (No.200801650002), the Natural Science Foundation of the Liaoning Scientific Committee
(No. 201102114), and Foundation of Liaoning Educational Committee (No. LT2011015).

\end{document}